\documentclass[nofootinbib,reprint]{revtex4-1}
\usepackage{subcaption}
\usepackage{graphicx}
\usepackage{amsmath}
\usepackage{amssymb}
\usepackage{amsthm}
\usepackage{bbm}             
\usepackage{bm}              
\usepackage{latexsym}
\usepackage[pdftex]{hyperref}
\usepackage[utf8]{inputenc}  
\usepackage{epsfig}

\usepackage{pstricks}
\usepackage{leftidx}        
\usepackage{graphicx}
\usepackage{psfrag}
\usepackage{rotating}
\usepackage{epsfig}
\usepackage{rotating}

\definecolor{Black}{rgb}{0.,0.,0.}
\definecolor{Blue}{rgb}{0.,0.,1.}
\definecolor{Green}{rgb}{0.,1.,0.}
\definecolor{Cyan}{rgb}{0.,1.,1.}
\definecolor{Red}{rgb}{1.,0.,0.}
\definecolor{Magenta}{rgb}{1.,0.,1.}
\definecolor{Yellow}{rgb}{1.,1.,0.}
\definecolor{White}{rgb}{1.,1.,1.}
\definecolor{Blue4}{rgb}{0.,0.,0.5625}
\definecolor{Blue3}{rgb}{0.,0.,0.6875}
\definecolor{Blue2}{rgb}{0.,0.,0.8125}
\definecolor{LtBlue}{rgb}{0.52734375,0.8046875,1.}
\definecolor{Green4}{rgb}{0.,0.5625,0.}
\definecolor{Green3}{rgb}{0.,0.6875,0.}
\definecolor{Green2}{rgb}{0.,0.8125,0.}
\definecolor{Cyan4}{rgb}{0.,0.5625,0.5625}
\definecolor{Cyan3}{rgb}{0.,0.6875,0.6875}
\definecolor{Cyan2}{rgb}{0.,0.8125,0.8125}
\definecolor{Red4}{rgb}{0.5625,0.,0.}
\definecolor{Red3}{rgb}{0.6875,0.,0.}
\definecolor{Red2}{rgb}{0.8125,0.,0.}
\definecolor{Magenta4}{rgb}{0.5625,0.,0.5625}
\definecolor{Magenta3}{rgb}{0.6875,0.,0.6875}
\definecolor{Magenta2}{rgb}{0.8125,0.,0.8125}
\definecolor{Brown4}{rgb}{0.5,0.1875,0.}
\definecolor{Brown3}{rgb}{0.625,0.25,0.}
\definecolor{Brown2}{rgb}{0.75,0.375,0.}
\definecolor{Pink4}{rgb}{1.,0.5,0.5}
\definecolor{Pink3}{rgb}{1.,0.625,0.625}
\definecolor{Pink2}{rgb}{1.,0.75,0.75}
\definecolor{Pink}{rgb}{1.,0.875,0.875}
\definecolor{Gold}{rgb}{1.,0.83984375,0.}

\definecolor{background}{cmyk}{0,0,0.3,0}
\definecolor{dgreen}{rgb}{0,.4,0}
\newrgbcolor{plum}{.7 .2 .7}
\newrgbcolor{darkgreen}{.0 .6 .0}
\newrgbcolor{peru}{.80 .52 .25}

\newcommand{\ist}[1]{\overset{\footnotesize(\ref{#1})}{=}}

\begin{document}

\title{\bf Charges and Electromagnetic radiation as topological excitations}
\author{Manfried Faber}

\affiliation{Atominstitut, Technische Universität Wien, Österreich}

\email{faber@kph.tuwien.ac.at}

\date{\today}

\begin{abstract}
We discuss a model with stable topological solitons in Minkowski space with only three degrees of freedom, the rotational angles of a spatial Dreibein. This model has four types of solitons differing in two topological quantum numbers which we identify with electric charge and spin. The vacuum has a two-dimensional degeneracy leading to two types of massless excitations, characterised by a topological quantum number which could have a physical equivalent in the photon number.
\end{abstract}



\maketitle

\section{Introduction}
In our mathematical description of nature we use two different concepts. Following Einstein we formulate gravity in a geometrical language, whereas particle physics uses the algebraic formulae of quantum field theory. To unify the two very successful theories most physicists try a quantise gravity. I go here into the opposite direction and follow first steps in a geometrical formulation of particle physics. In which direction to go in our investigations we should find out from experiments. Nature may give us some hints about the mechanisms. My first intuition I get from the sine-Gordon model and its experimental realisation with a pendulum model. The mathematics of the sine-Gordon model is nicely described in \cite{Remoissenet:1999wa}. Here I want only to repeat the most interesting physical pictures of this model. It is a fully relativistic model in 1+1D, where the velocity of light corresponds to the propagation velocity $c$ of small amplitude waves. In the experimental realisation, see Ref.~\cite{Remoissenet:1999wa} this velocity is of the order of 1~m/s. Besides waves we find two types of particle-like excitations, kinks and anti-kinks. They behave in many ways like particles. Their energy density is concentrated in a certain region in space with a total energy defining the mass. The three contributions to the energy, stress energy, potential energy and kinetic energy have different dependencies on the velocity $v$ of a moving kink. The potential energy is decreasing, stress energy and kinetic energy are increasing with $v$. As expected for a relativistic model the three contributions nicely sum up to a mass increasing with $\gamma=1/\sqrt{1-(v/c)^2}$. A moving kink is Lorentz contracted. In the mechanical model we can easily imagine how nature does to decrease the size of the kink. To accelerate the pendula at the front of the kink the angle between them has to increase leading to a smaller size of the kink. The pendulum model gives me some idea how nature could work to realise the phenomena of special relativity. Even more impressive is the existence of two types of kinks, kinks and antikinks and their interaction. They behave like charged particles, kinks and kinks repel, kinks and antikinks attract each other. In soft collisions kinks behave similar to billiard balls. In hard collisions the diameters of kinks shrink proportional to $1/\gamma$ and get point-like. Further, we can observe how kinks and antikinks annihilate. In the mechanical model the annihilation due to friction effects gives rise to the emission of waves. In the abstract theoretical model solitons and antisolitons get through each other with a small time-delay. Mathematically we can separate the various kink configurations in homotopy classes differing in their winding number. This is condensed in the relation $\Pi_1(\mathbbm S^1)=\mathbbm Z$ of homotopy theory.

A second hint I get from a simple model teaching us about the nature of $4\pi$-rotations. I saw it for the first time in Figure 41.6 on page 1149 of ``Gravitation'' by Misner, Thorne and Wheeler~\cite{misner1973gravitation}. A ball is attached with several wires to the surrounding, e.g. with eight wires to the corners of a cube. Rotating the ball one or two times around some axis leads to a complete mess of the strings. But after a $4\pi$ rotation one can disentangle the wires without moving the ball. We can learn from this model that a body which is connected to the surrounding returns only after a $4\pi$-rotation to his original state. For a disconnected body this happens already after a $2\pi$-rotation. This is mathematically formulated in the relation $\Pi_1(\mathbbm S^3)=1$. There is a continous transition between $4\pi$-rotations and no rotation. This ball model gives me a hint how possibly nature realises particles with spin 1/2 just by connecting them with the surrounding.

Thirdly, I want to mention that observing phenomena at and below the atomic scale we always observe particles or clicks and never waves. Remember the double-slit experiment. In this interference experiment of electrons or photons the wave-picture appears only after several hundreds or thousands of particles have been registered on the screen.

In the main part of this article I describe a Lorentz covariant model which has stable topological exciations with properties of particles. In a certain sense it is a generalisation of the sine-Gordon model to 3+1D. Several features of this model were already described in a few articles~\cite{Faber:1999ia,Faber:2002nw,Borisyuk:2007bd,Faber:2008hr,Faber:2012zz,Faber:2014bxa}. In this article I will mainly concentrate on topological questions.

\section{Definition of the model}
We are using a scalar SO(3)-field in 3+1D Minkowski space. The only degrees of freedom of this model are therefore three rotational angles, e.g. the three Euler angles, describing the rotations of a spatial Dreibein. Simpler than to use SO(3)-matrices is to work with SU(2), i.e. with $2\times2$-matrices. Since SU(2) is the double covering group of SO(3) there is an essential difference betweeen SU(2)- and SO(3)-fields. Every field configuration of an SO(3)-field is twice realised by SU(2)-fields. The two realisation differ by a transformation with the non-trivial center element, by a $2\pi$ rotation. This property we have to remember using SU(2)-matrices
\begin{equation}\label{DefQ}
Q(x)=\mathrm e^{-\mathrm i\alpha(x)\vec\sigma\vec n(x)}
=\cos\alpha(x)-\mathrm i\vec\sigma\vec n(x)\sin\alpha(x).
\end{equation}
at every site $x$ in $M_4$. The symbol $Q$ we are using reminds to quaternions, invented by Benjamin Olinde Rodrigues~\cite{Rodrigues:1840aa} in the year 1840, to describe active rotations with the three imaginary quaternionic units $\mathbf i,\mathbf j,\mathbf k$. In Eq.~(\ref{DefQ}) they are represented by Pauli matrices $\mathbf i:=-\mathrm i\sigma_1,\mathbf j:=-\mathrm i\sigma_2,\mathbf k:=-\mathrm i\sigma_3$. $\vec n$ is a three component unit vector and $\vec\sigma\vec n:=\sum_{i=1}^3\sigma_in_i$ is the component of the Pauli matrices in direction of $\vec n$. Rotations are unit quaternions $Q=q_0-\mathrm i\vec\sigma\vec q$ with $q_0^2+\vec q^2=1$. Their manifold is isomorphic to $\mathbbm S^3$.

The idea for the definition of the dynamical part of the Lagrangian is its identification with the square of the area density on $\mathbbm S^3$ in appropriate units of an action density. We start defining tangential vectors
\begin{equation}\label{DefTangVek}
\partial_\mu Q:=-\mathrm i\vec\sigma\vec\Gamma_\mu Q\quad\textrm{with}\quad
\vec\sigma\vec\Gamma_\mu:=\sum_{i=1}^3\sigma_i\Gamma_{\mu i},
\end{equation}
(tangential one-forms) to $\mathbbm S^3$. We would like to emphasise that $\vec A_\mu=2\vec\Gamma_\mu$ is a trivial connection but $\vec\Gamma_\mu$ is not. With the cross-product $\vec R_{\mu\nu}:=\vec\Gamma_\mu\times\vec\Gamma_\nu$ we can get the square of the area density $\vec R_{\mu\nu}\vec R^{\mu\nu}$ and define the Lagrangian in appropriate SI-units with $\alpha_f:=\frac{e_0^2}{4\pi\epsilon_0\hbar c}$
\begin{equation}\begin{aligned}\label{DefL}
\mathcal L:=&\mathcal L_\mathrm{dyn}-\mathcal H_\mathrm{pot}
:=-\frac{\alpha_f\hbar c}{4\pi}\left(\frac{1}{4}\,\vec R_{\mu\nu}\vec R^{\mu\nu}
+\Lambda(q_0)\right)\\&\vec R_{\mu\nu}:=
\vec\Gamma_\mu\times\vec\Gamma_\nu,\quad\vec\Gamma_\mu\ist{DefTangVek}
\frac{\mathrm i}{2}\mathrm{Sp}(\vec\sigma\;\partial_\mu QQ^\dagger).
\end{aligned}\end{equation}
Up to a proportionality factor, the kinetic term $\mathcal L_\mathrm{dyn}$ of this model agrees with the Skyrme term in the Skyrme model~\cite{Skyrme:1961vq}. To get stable solitons, Skyrme suggested to use a mass term for the vector field $\vec\Gamma_\mu$ leading to Skyrmions with short-range forces which are accepted as approximations for nucleons. We want to describe particles with long-range Coulombic forces and have therefore to avoid the Skyrme term. The Hobart-Derrick theorem~\cite{Hobart:1963rh,Derrick:1964gh} allows as additional terms only terms without derivative, a potential term, which we chose as
\begin{equation}\label{DefPot}
\Lambda(q_0)=q_0^{2m}.
\end{equation}
Therefore, we have a two-dimensional manifold of degenerate vacua, the equatorial sphere $\mathbbm S^2_\mathrm{equ}$ defined by $q_0=0$. The choice of the potential term~(\ref{DefPot}) has two immediate physical consequences. Two Goldstone bosons, which we would like to identify with the two photon degrees of freedom and non-trivial field configurations of finite energy which can be classified by $\Pi_2(\mathbbm S^2_\mathrm{equ})=\mathbbm Z$, by the map of $\mathbbm S^2_\mathrm{equ}$ to the sphere $\mathbbm S^2_\infty$ at spatial infinity.

\section{Stable solitons of finite energy}
Inserting the time-independent hedgehog-ansatz
\begin{equation}\begin{aligned}\label{RegularIgel}
&q_0=\cos\alpha(r),\quad\vec q=\vec n({\mathbf r})\sin\alpha(r),\\
&\vec n({\mathbf r})=\frac{\vec r}{r},\quad\alpha(r)\in[0,\frac{\pi}{2}]
\end{aligned}\end{equation}
into the lagrangian~(\ref{DefL}) we get the Euler-Lagrange equation
\begin{equation}\begin{aligned}\label{nlDE}
&\partial^2_\rho\cos\alpha+\frac{(1-\cos^2\alpha)\cos\alpha}{\rho^2}
 -m\rho^2\cos^{2m-1}\alpha=0\\&\textrm{with}\quad\rho=\frac{r}{r_0},
\end{aligned}\end{equation}
a non-linear differential equation which we can most easily solve for $m=3$. It has the simple solution
\begin{equation}\begin{aligned}\label{MinLoesm3}
&\alpha(r)=\arctan\rho,\\&\sin\alpha(r)=\frac\rho{\sqrt{1+\rho^2}},
\qquad\cos\alpha(r)=\frac{1}{\sqrt{1+\rho^2}}
\end{aligned}\end{equation}
leading to the radial energy density
\begin{equation}\label{radialdensities}
h=\frac{\alpha_f\hbar c}{r_0}\Big[\frac{\rho^2}{2(1+\rho^2)^2}
 +\frac{\rho^2}{(1+\rho^2)^3}+\frac{\rho^2}{(1+\rho^2)^3}\Big].
\end{equation}
\begin{figure}[h!]
\centering\scalebox{1.0}{\input{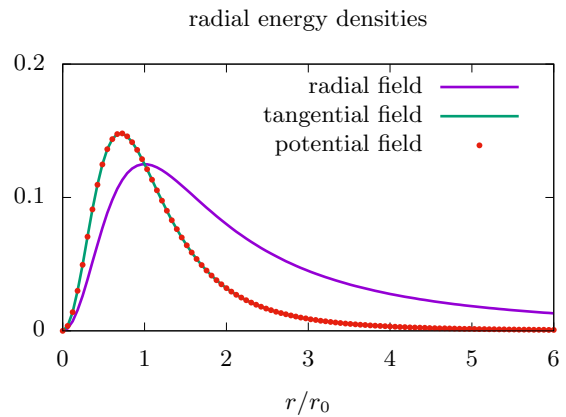}}
\caption{Contributions to the radial energy density according to Eq.~(\ref{radialdensities}) in units of $\frac{\alpha_f\hbar c}{r_0}$.}\label{enedicht}
\end{figure}
The radial dependence of its three contributions are shown in Fig.~\ref{enedicht}. At large distances the radial energy density agrees with the Coulombic energy density of a point charge. As one can clearly see, the singularity at the origin, point-charges usually are suffering from, is removed. The tangential and the potential energy density have equal shapes and decay faster. They lead to a deviation from the Coulomb interaction at distances in the order of $r_0$ and smaller. The total energy sums up to $E=\frac{\alpha_f\hbar c}{r_0}\frac{\pi}{4}$. Comparing this result with the rest energy of an electron we get a value for the scale $r_0$ of 2.21~fm.

\begin{figure}[h!]
\centering\includegraphics[scale=0.20]{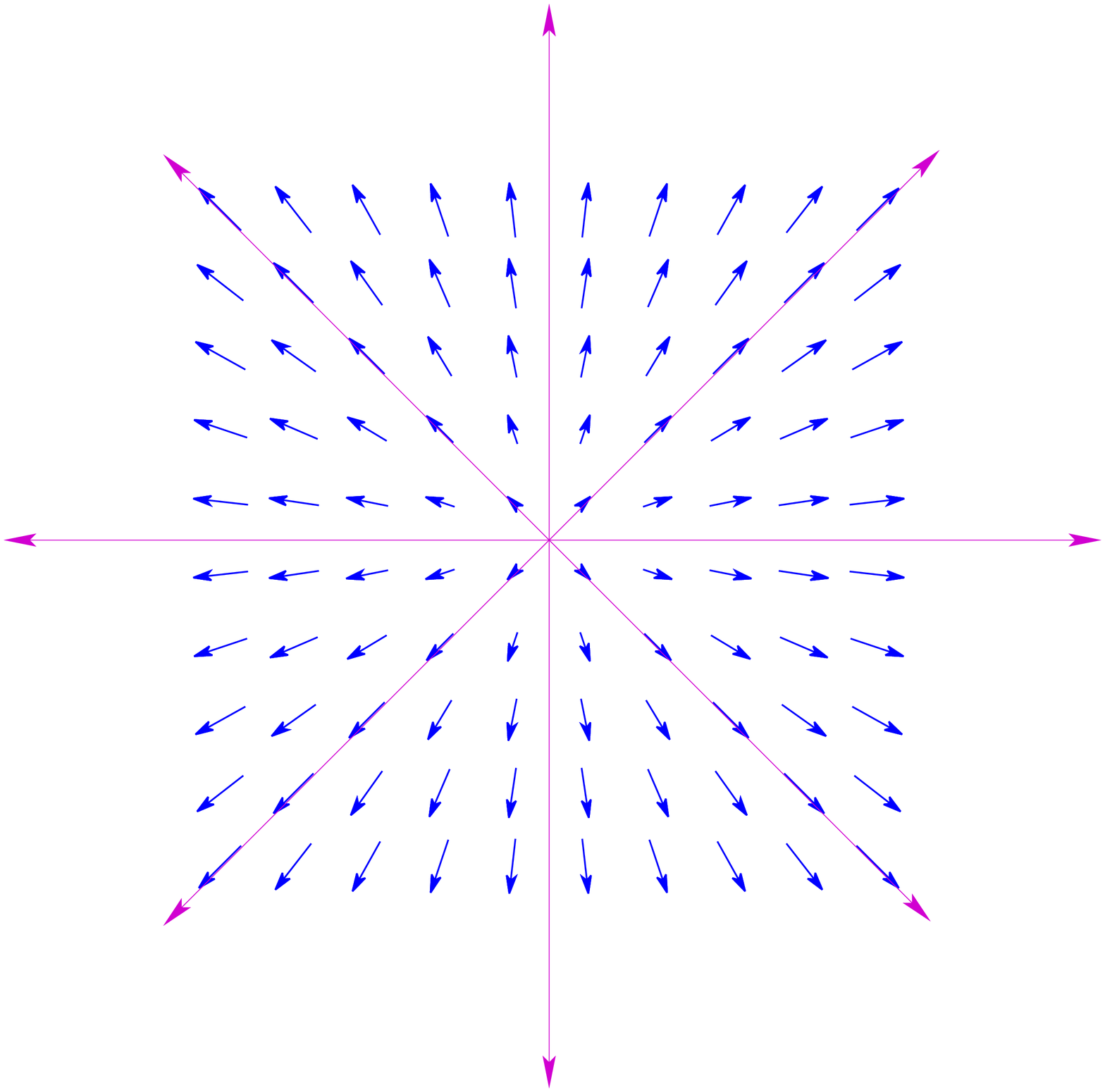}\\
\scalebox{0.4}{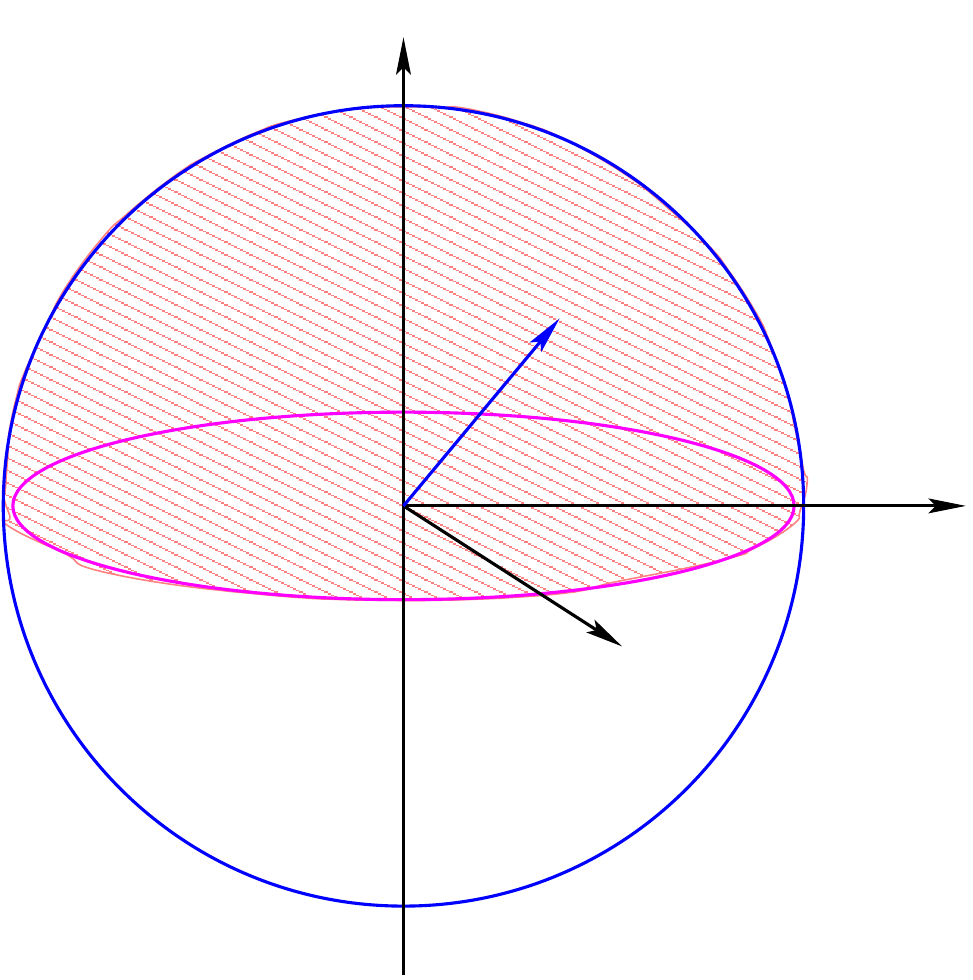}
\caption{Above: Scheme of the hedge-hog configuration~(\ref{RegularIgel}). The small blue arrows show the imaginary part $\vec q=\vec n({\mathbf r})\sin\alpha(r)$ of the $Q$-field in an arbitrary plane through the origin. The long magenta lines indicate the electric field lines of a point-charge, the lines of constant $\vec n$-field. Below: Half-sphere covered by the hedge-hog configuration.}
\label{schemaelek}
\end{figure}

The hedge-hog configuration, defined in Eq.~(\ref{RegularIgel}), is schematically depicted in the upper diagram of Fig.~\ref{schemaelek}. It maps $\mathbbm R^3$ to half of $\mathbbm S^3$. This half-sphere is indicated in the lower diagram of Fig.~\ref{schemaelek}.
 By the electric field the hedge-hog is wired to the surrounding. If the centre of the hedge-hog is rotated by $4\pi$, the tangled lines of constant $\vec n$-field can be disentangled without further rotation of the centre, and the original configuration can get restored. This behaviour reminds of the rotational property of spin-1/2 particles.

\section{Topological quantum numbers}
A further relation to spin we can find in the number of coverings of $\mathbbm S^3$, the topological charge $\mathcal Q$, which we define in spherical coordinates $r,\vartheta,\varphi$ by
\begin{equation}\label{TopLad}
\mathcal Q:=\frac{1}{2\pi^2}
\int_0^\infty\mathrm dr\int_0^\pi\mathrm d\vartheta\int_0^{2\pi}\mathrm d\varphi\,
\vec\Gamma_r(\vec\Gamma_\vartheta\times\vec\Gamma_\varphi).
\end{equation}
\begin{table}[h!]\vspace{2mm}
\begin{center}\hspace*{-2mm}\begin{tabular}{|c||c|c|c|c|}\hline 
Transf.&$1$&$z$&$\Pi_n$&$z\Pi_n$\\
\hline\hline
$\vec n$&$\vec r/r$&$-\vec r/r$&-$\vec r/r$&$\vec r/r$\\\hline
$q_0$&$\ge 0$&$\le 0$&$\ge 0$&$\le 0$\\\hline\hline
$Q_\mathrm{el}$&$-1$&$1$&$1$&$-1$\\\hline
$\mathcal Q$&$1/2$&$1/2$&$-1/2$&$-1/2$\\\hline
diagram&\includegraphics[scale=0.1]{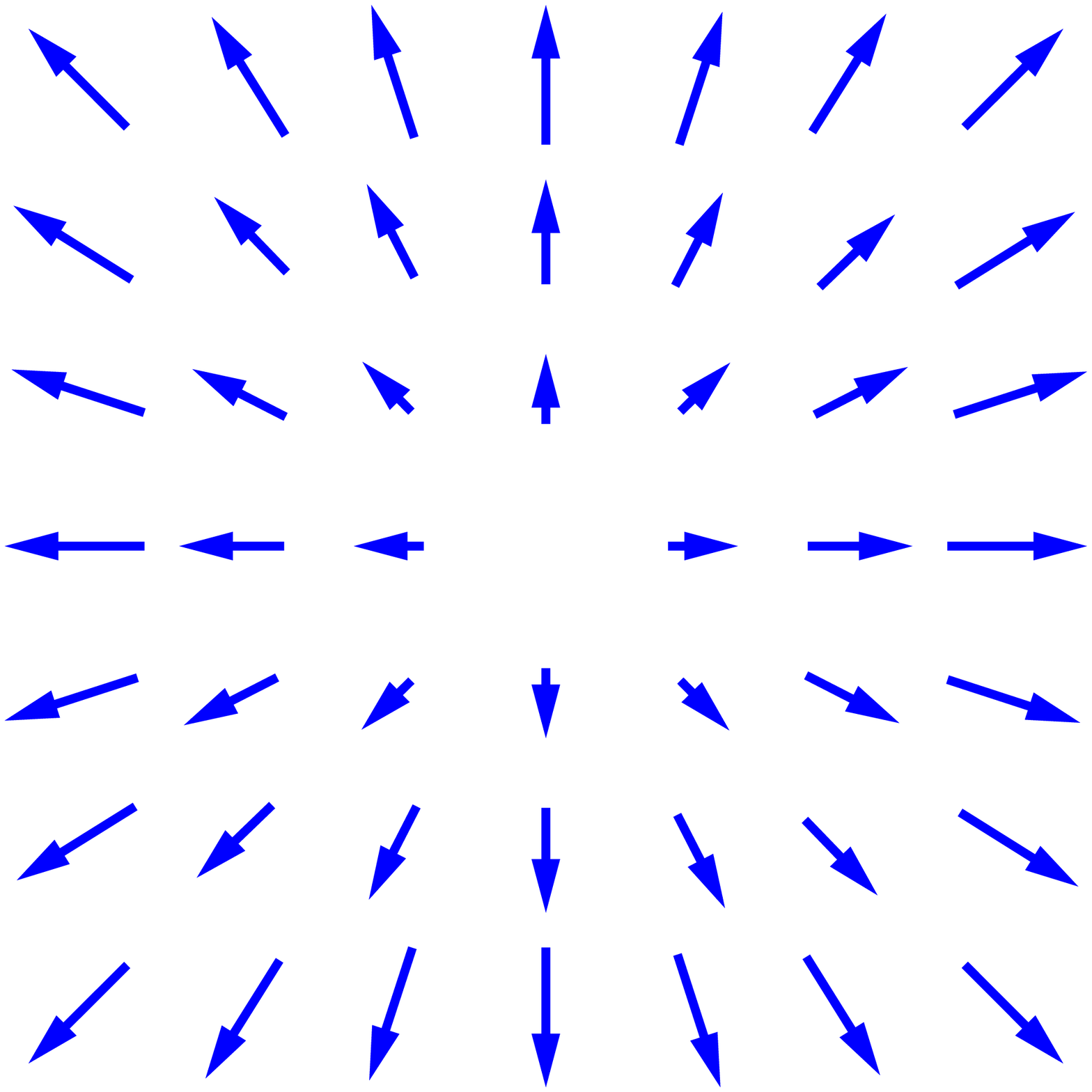}&\includegraphics[scale=0.1]{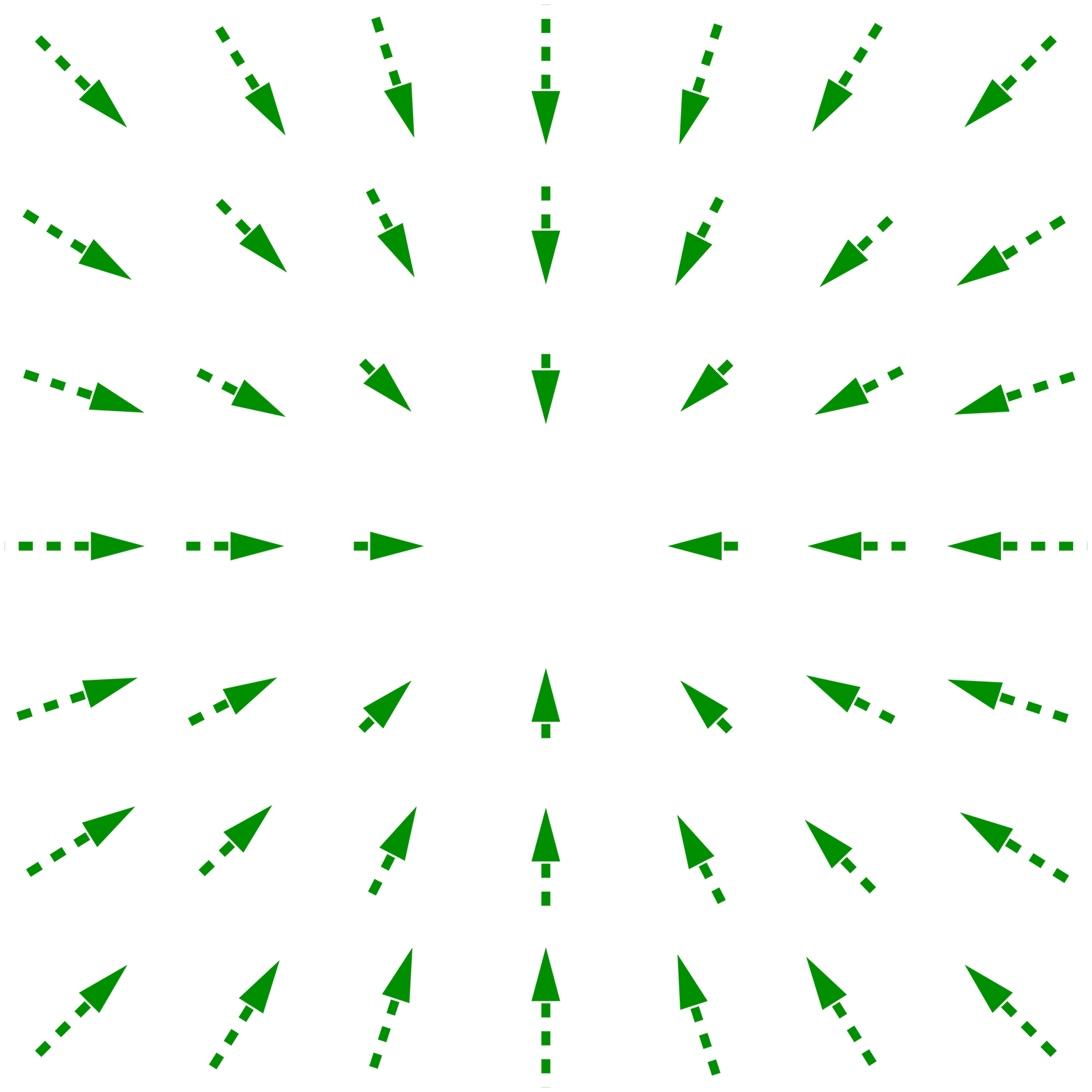}&\includegraphics[scale=0.1]{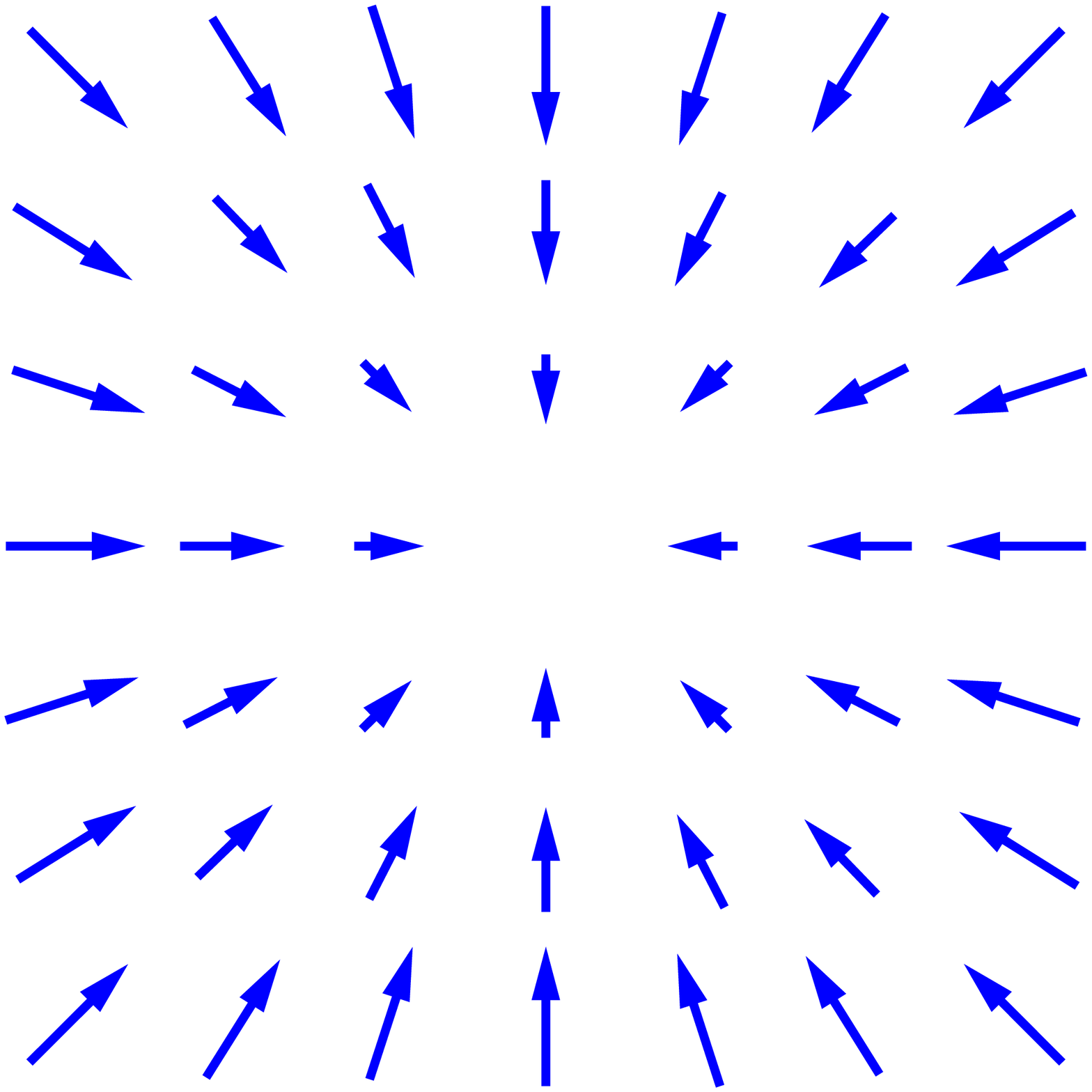}&\includegraphics[scale=0.1]{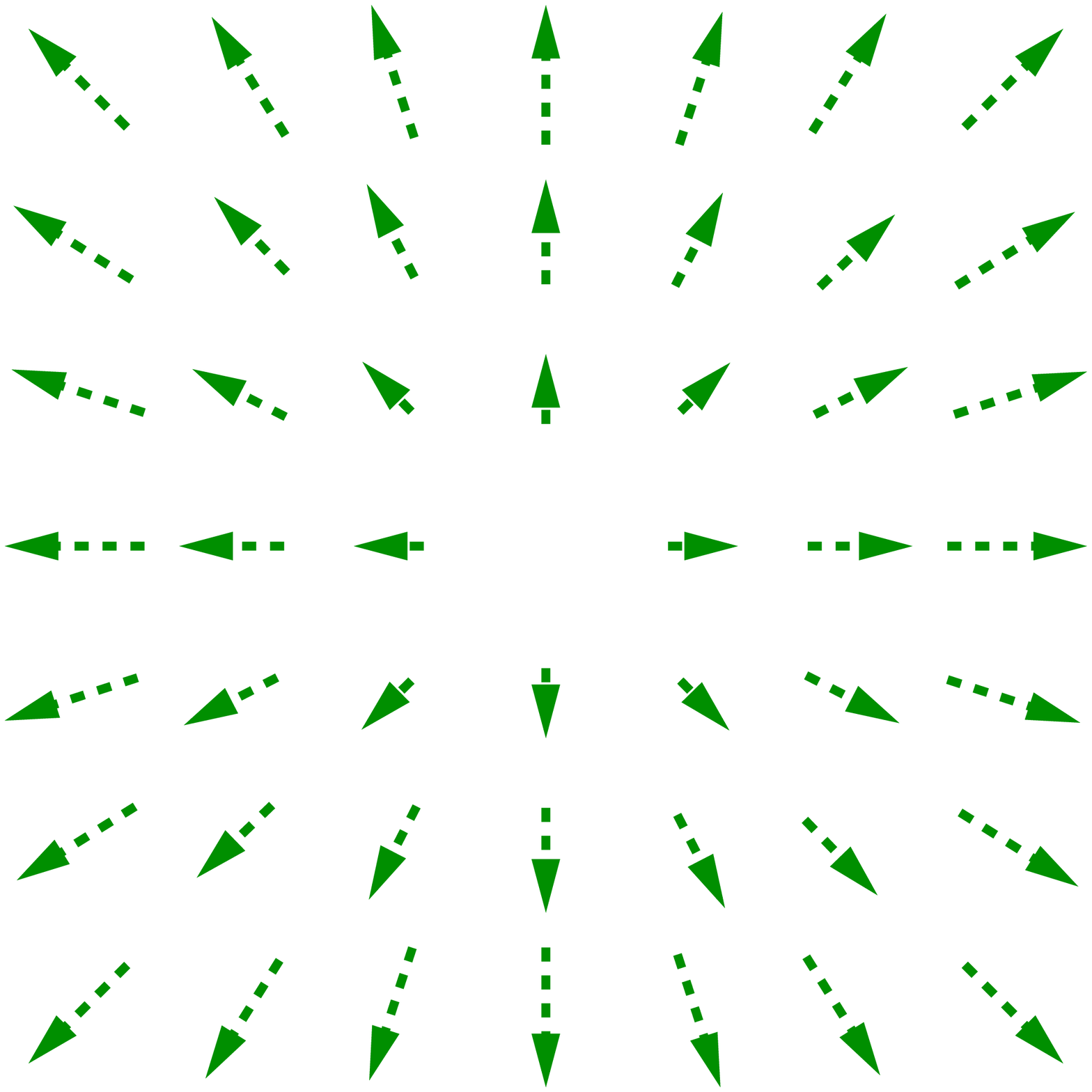}\\\hline
\end{tabular}\end{center}\caption{Soliton types differ by transformations with $\Pi_n$ and $z$. The signs of $\vec n$ and $q_0$ and the topological quantum numbers $Q_\mathrm{el}$ and $\mathcal Q$ are quoted. The configurations are spherical symmetric.}\label{SolTypes}\end{table}
The configuration of Eq.~(\ref{RegularIgel}) and Fig.~\ref{schemaelek} results in $\mathcal Q=1/2$. Continuous modifications of the soliton field do not change $\mathcal Q$ and the homotopy class of the configuration.  The homotopy class can be changed by parity transformations $\Pi_n:\vec n\mapsto-\vec n$ in the internal space and by transformations with the non-trivial centre element $z=-1$. The four types of solitons which we get by these transformations are shown in Tab.~\ref{SolTypes}. $Q$-values with $q_0\ge0$ are indicated by full blue arrows and with $q_0\le0$ by dashed green arrows. Besides the topological charge, the four configurations differ in the direction of the rotational axis of the $Q$-field at infinity which we will relate in Eq.~(\ref{EDynGrenz}-\ref{EdynGauss}) with the electric charge $Q_\mathrm{el}$. This field at large distances from the centre determines the interaction with other solitons, attraction or repulsion. The pairs of configurations with the same electric charge differ in the value $\pm\frac{1}{2}$ of the topological charge, in the chirality. We can combine pairs of solitons with $\mathcal Q=\pm\frac{1}{2}$ either to $\mathcal Q=0$ or $\mathcal Q=\pm1$. Since SU(2) is the spin group and due to the above described properties of the soliton configurations under $4\pi$ rotations we dare to identify the absolute value of the topological charge with the spin quantum number
\begin{equation}\label{SpinAssign}
s=|\mathcal Q|.
\end{equation}
We would like to remind that within SO(3) the two configurations which differ in the sign of $\mathcal Q$ are identical.

Traversing the centre of a soliton we follow a rotation of the local Dreibeins by $\pm2\pi$. This may answer a question, posed by Tsung-Dao Lee in a talk given in Vienna in the 1980's: ``Why does the mass break chiral symmetry?''

\begin{figure}[h!]
\centering\includegraphics[scale=0.3]{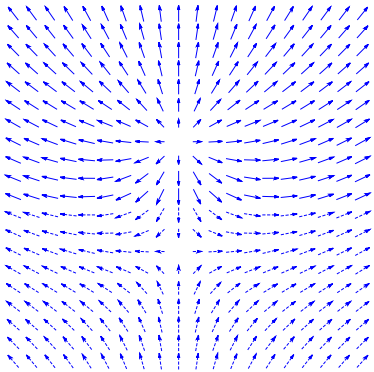}
\caption{Dipole configuration with $S=1$. The small blue arrows show the imaginary part $\vec q=\vec n({\mathbf r})\sin\alpha(r)$ of the $Q$-field in an arbitrary plane through the centres of both solitons. Full arrows correspond to $q_0\le0$ and dotted arrows to $q_0\le0$.}\label{Dipole1}
\end{figure}

\begin{figure}[h!]
\centering\includegraphics[scale=0.3]{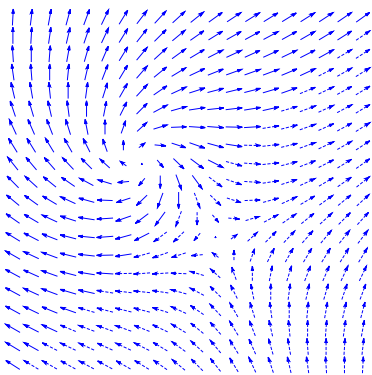}
\caption{Configuration of Fig.~\ref{Dipole1}, after rotation by $\frac{\pi}{4}$.}\label{Dipole45}
\end{figure}

Besides its group theoretical properties, spin is a contribution to the total angular momentum. We will investigate these angular momentum properties in a dipole configuration. Since a dipole is uncharged, the field at infinity is independent of the direction, it approaches e.g. $\lim_{r\to\infty}Q(\mathbf r)=-\mathrm i\sigma_3$. The symmetry of the vacuum is broken. We can combine the first configuration in Table~\ref{SolTypes} with the second to total spin $S=1$ or with the third to $S=0$. The energy for $S=0$ is slightly lower than that for $S=1$, the configuration shown in Fig.~\ref{Dipole1}. During a rotation of the dipole the vacuum has to remain unchanged, a rigid rotation is not possible. If the dipole axis is rotated, e.g. by $\frac{\pi}{4}$, as shown in Fig.~\ref{Dipole45}, the centres of the solitons have to rotate by the same angle. These rotations contributes to the total angular momentum.

Dipole configurations are not stable due to the attractive interaction between the charges. Their time evolution follows the general equations of motion~\cite{Faber:1999ia}
\begin{equation}\label{BGl1}
\partial_\mu[\vec\Gamma_\nu\times\vec R^{\mu\nu}]
 +\vec q\,\frac{\mathrm d\Lambda}{\mathrm d q_0}=0
\end{equation}
derived from the Lagrangian~(\ref{DefL}). The energy-momentum tensor
\begin{equation}
{\Theta^\mu}_\nu=
 -\frac{\alpha_f\hbar c}{4\pi}
 \left\{\left(\vec\Gamma_\nu\times\vec\Gamma_\sigma\right)
 \left(\vec\Gamma^\mu\times\vec\Gamma^\sigma\right)\right\}-
 \mathcal L(x)\,\delta^\mu_\nu .
\label{ExpressionEMT}
\end{equation}
turns out to be symmetric. There is no special symmetrisation necessary as in Maxwell's elecrodynamics~\cite{jackson_classical_1999}. Since there are no external forces, the force density is vanishing
\begin{equation}\label{verschwKraft}
f_\nu=\partial_\mu\Theta^\mu_{\;\nu}=0.
\end{equation}

To determine a static potential one has to fix the positions of the centres at a chosen distance $d$. For distances $d\gg r_0$ the Coulombic behaviour is nicely reproduced, as can be seen in the diploma theses~\cite{Wabnig2001,Resch2011,Theuerkauf2016}. For distances $d\overset{<}{\approx}r_0$ the interaction strength increases.

\section{Electrodynamic limit}
A pure Coulombic behaviour one gets in the limit $r_0\to0$ where one arrives at the  Wu-Yang description~\cite{Wu:1975vq} of dual Dirac monopoles by two degrees of freedom which we can choose as a normalised three dimensional vector field $\vec n$. In this limit we get
\begin{equation}\begin{aligned}\label{EDynGrenz}
&Q(x)=-\mathrm i\vec\sigma\vec n(x),\\
&\vec\Gamma_\mu(x)\ist{DefTangVek}\vec n(x)\times\partial_\mu\vec n(x),\\
&\vec R_{\mu\nu}(x)\ist{DefL}\partial_\mu\vec n(x)\times\partial_\nu\vec n(x)
\end{aligned}\end{equation}
This is a description where the singularity of the Dirac string is removed, but the singularity of the Coulomb field is still present. The Lagrangian~(\ref{DefL}) reduces to\footnote{The same degrees of freedom but a different Lagrangian is used in the Fadeev-Niemi = Skyrme-Fadeev = Baby-Skyrme model~\cite{Faddeev:1975tz,Faddeev1976,Ferreira:2000hz}. The same Lagrangian was used in ~\cite{Ferreira:2006wg}.}
\begin{equation}\begin{aligned}\label{EDynGrenz}
\mathcal L_\mathrm{ED}=-\frac{1}{4\mu_0}\leftidx{^*}F_{\mu\nu}(x)\leftidx{^*}F^{\mu\nu}(x)
\end{aligned}\end{equation}
with the dual field strength tensor
\begin{equation}\begin{aligned}\label{EDynGrenz}
\leftidx{^*}F_{\mu\nu}(x)
&=-\frac{e_0}{4\pi\varepsilon_0 c}\vec R_{\mu\nu}\vec n=\\
&=-\frac{e_0}{4\pi\varepsilon_0 c}
\vec n(x)[\partial_\mu\vec n(x)\times\partial_\nu\vec n(x)].
\end{aligned}\end{equation}
In this limit hedge-hogs are characterised by point-like singularities in space and closed world-lines of line-like singularities in space-time
\begin{equation}\label{EdynGrenJ}
j^\mu=-e_0 c\sum_{i=1}^N\int\mathrm d\tau_i\frac{\mathrm dX^\mu(\tau_i)}
{\mathrm d\tau_i}\,\delta^4(x-X(\tau_i))=(c\rho,\mathbf{j}).
\end{equation}
Charges and fields are related by the inhomogeneous Maxwell-equations
\begin{equation}\label{EdynGauss}
\frac{1}{2\mu_0}\oint_{\partial V}\mathrm dx^\mu\mathrm dx^\nu{^*F}_{\mu\nu}
 =\frac{1}{6}\int_V\mathrm dx^\mu\mathrm dx^\nu\mathrm dx^\rho
 \epsilon_{\mu\nu\rho\sigma}j^\sigma.
\end{equation}
Already here we see an essential difference to Maxwell's theory. Charges are quantized, there are no other charges possible than integer multiples of the elementary charge $e_0$. Two further differences we get from the equations of motion
\begin{equation}\label{BewGlEdyn}
\partial_\mu\vec n\,g^\mu=0.
\end{equation}
They allow for non-vanishing magnetic currents
\begin{equation}\label{magcur}
g^\mu=c\,\partial_\nu\hspace{0.2mm}{^\star}\hspace{-0.2mm}f^{\nu\mu}
\quad\Leftrightarrow\quad
\begin{cases}
&\rho_\mathrm{mag}=\nabla\mathbf B,\\
&\mathbf g=-\nabla\times\mathbf E-\partial_t\mathbf B.
\end{cases}
\end{equation}
The solutions of the homogeneous Maxwell equations fulfil the equations of motion~(\ref{BewGlEdyn}). But there are further solutions possible which fulfil the relations
\begin{equation}
\mathbf B\,\mathbf g\ist{BewGlEdyn}0,\quad
c^2\mathbf B\,\rho_{\rm mag}\ist{BewGlEdyn}\mathbf g\times\mathbf E,
\end{equation}
equivalent to the equations of motion~(\ref{BewGlEdyn}). The presence of unquantized closed magnetic currents may be a discrepancy to experiments. But we could explain it with the observation, mentioned in the introduction, that in the experiment we only detect particles and never waves. We can speculate that such currents contribute to the recently intensively discussed dark matter. Further, we read from these equations that there are no solutions possible where $\mathbf E$ and $\mathbf B$ are parallel. This seems obviously also in contradiction to experiments, where it is rather simple to produce static parallel electric and magnetic fields. In this case it is more difficult to find an excuse. It could be that $\mathbf E$ and $\mathbf B$ are locally perpendicular and they appear to be parallel only in the average over space or time. This is the price to pay for restricting charges to integer multiples to the elementary charge and the fields to the two degrees of freedom of the $\vec n$-field.

\section{Coulomb and Lorentz forces}
By the artificial splitting~(\ref{EDynGrenz}) of a single field $Q(x)$ in particles and their fields we reduce ${\Theta^\mu}_\nu$ to the symmetric energy-momentum tensor
\begin{equation}\label{ExpEMomTen}
{T^\mu}_\nu(x)\ist{ExpressionEMT}
-\frac{1}{\mu_0}{\hspace{0.5mm}^\star}\hspace{-0.5mm}F_{\nu\sigma}(x)
 {\hspace{0.5mm}^\star}\hspace{-0.5mm}F^{\mu\sigma}(x)-\mathcal L_{\rm ED}(x)\,\delta^\mu_\nu
\end{equation}
and split the force density in two contributions 
\begin{equation}\label{totforczero}
f_\nu=\partial_\mu{\Theta^\mu}_\nu
=f^\mu_\mathrm{charges}+\partial^\nu{T^\mu}_\nu\ist{verschwKraft}0
\end{equation}
showing clearly the presence of Coulomb and Lorentz forces
\begin{eqnarray}\label{forcedensity}
f^0_\mathrm{charges}=\frac{1}{c}\,\mathbf j\,\mathbf E,\quad
\mathbf{f}_\mathrm{charges}=\rho\,\mathbf E+\mathbf j\times\mathbf B.
\end{eqnarray}
Here we would like to underline, that the magnetic currents $\mathbf g$ don't contribute to electromagnetic forces.

\section{U(1) gauge invariance}
A U(1) gauge invariance appears as a rotational invariance by $\omega(x)$ around the $\vec n$-axis. By a rotation in colour space with
\begin{equation}\label{gotoabel}
\Omega(x)=\mathrm e^{\mathrm i\theta(x)\vec e_\phi(x)\vec L}
\mathrm e^{\mathrm i\omega(x)\vec n\vec L}
\end{equation}
we can rotate the $\vec n$-field in 3-direction
\begin{equation}\label{nTrafo}
\vec n=\begin{pmatrix}\sin\theta\cos\phi\\\sin\theta\sin\phi\\\cos\theta
\end{pmatrix}\to\vec n^\prime:=\Omega\vec n=
\left(\begin{array}{c}0\\0\\1\\\end{array}\right)=\vec e_3.
\end{equation}
Under this transformation the vector field $\vec L\,\vec\Gamma_\mu$ transforms to $\vec L\,\vec\Gamma_\mu^\prime=
\Omega(\Gamma_\mu-\mathrm i\partial_\mu)\,\Omega^\dagger$ with
\begin{equation}\label{Gammaabel}
\vec\Gamma_\mu^\prime=[(1-\cos\theta)\partial_\mu\phi
+\partial_\mu\omega]\;\vec e_3.
\end{equation}
The curvature $\vec R_{\mu\nu}$ gets rotor form and turns out to be invariant against the rotations with $\omega(x)$
\begin{equation}\begin{aligned}\label{Rabel}
\vec R_{\mu\nu}
&=\partial_\mu\vec\Gamma_\nu^\prime-\partial_\nu\vec\Gamma_\mu^\prime=\\
&=[-\partial_\mu\cos\theta\,\partial_\nu\phi
+\partial_\nu\cos\theta\,\partial_\mu\phi]\vec e_3.
\end{aligned}\end{equation}
The far-field of a hedge-hog soliton is the electric field strength of a classical electron and the vector field of a dual Dirac monopole
\begin{equation}
\vec E_{r}^\prime=\frac{e_0}{4\pi\varepsilon_0}\frac{\vec e_3}{r^2},\qquad
\frac{\vec\Gamma_\varphi(\vartheta)}{r\sin\vartheta}=\frac{1-\cos\vartheta}{r\sin\vartheta}\vec e_3
\end{equation}
with a Dirac string along the 3-direction.

\section{Hopf number}
In Fig.~\ref{schemaelek}a we realised that the field lines of a point charge are lines of constant $\vec n$-field. This relation we find also for the dipole-fields in Fig.~\ref{Dipole1} and \ref{Dipole45}. Since in our model the vacuum has broken symmetry the field at infinity ``$\infty$'' is independent of the direction. Thus we get the isomorphism $\mathbb R^3 \cup \infty \sim \mathcal{S}^3$. Due to the topological relation $\pi_3(\mathcal{S}^2)=\mathbb{Z}$ there is an additional quantum number for the $\vec n$-field, the Hopf number or Gauß linking number $v$ of fibres $\mathcal F$ defined by $\vec n_\mathcal{F}=$const, thus by certain values $\theta_\mathcal{F}$ and $\phi_\mathcal{F}$. This linking number is especially interesting in regions where we can neglect the influence of charges, in regions of pure $\vec n$-field, where the Lagrangian reduces to $\mathcal L_\mathrm{ED}$ of Eq.~(\ref {EDynGrenz}). According to the Hobart-Derrick theorem~\cite{Hobart:1963rh,Derrick:1964gh} such configurations are unstable, if they move with a velocity slower than the speed of light $c$. Moving with $c$ they have an action constant in time and can be used to describe electromagnetic waves. For simplicity we choose the z-axis in the direction of motion, then it is sufficient to describe these configurations in the $\mathbb R^3$ given by the coordinates $x$,$y$ and $\zeta=z-ct$. The linking number $v$ is a topological invariant. Its natural physical equivalent is the number of photons $n_\gamma$ in this configuration.

\begin{figure}[h!]
\includegraphics[scale=0.6]{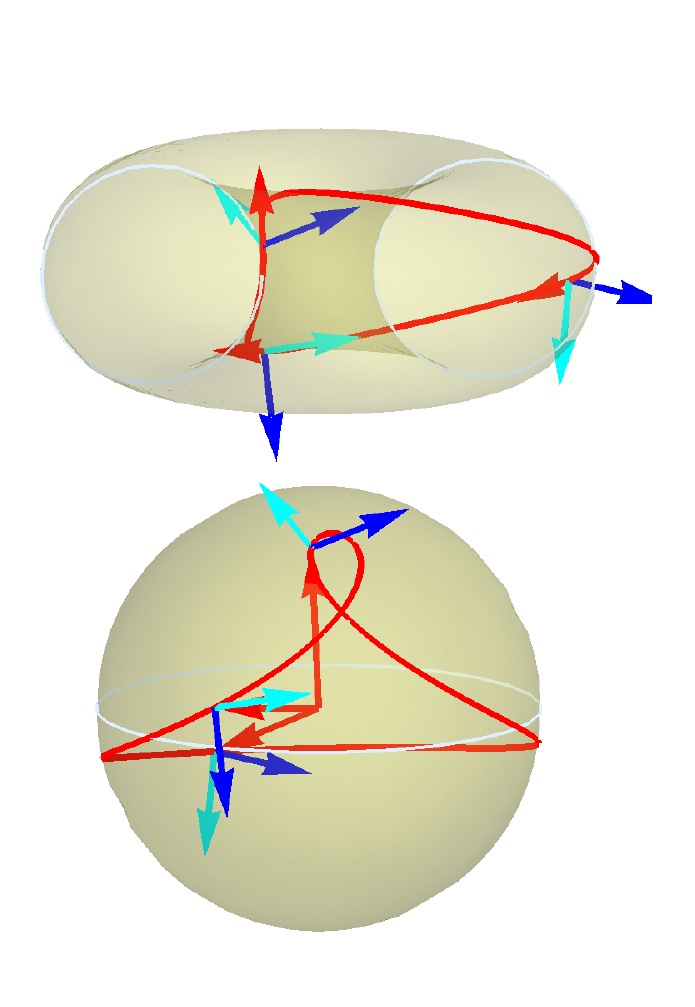}
\includegraphics[scale=0.6]{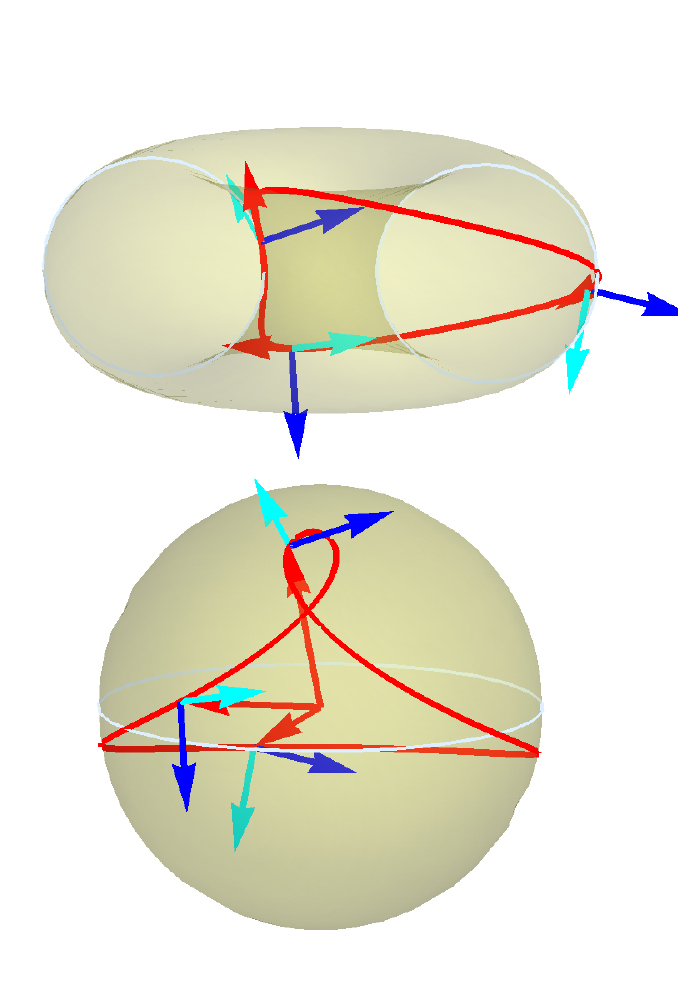}
\caption{Stereographic view with parallel eyes at a fibre neighbourhood. Above: Fibre $\mathcal{F}$ on a torus in $\mathbb R^3$. Dreibeins of red tangential vectors and blue and cyan tangential vectors are drawn at three positions. Below: Gauß map $\mathcal C_\mathcal{F}$ of $\mathcal F$ with red radial vector $\vec e_\mathcal{F}^{\,\prime}(s)$ and blue and cyan tangential vectors at the corresponding three positions.}
\label{DrehFaserUmg}
\end{figure}

\begin{figure}[h!]
\includegraphics[scale=0.45]{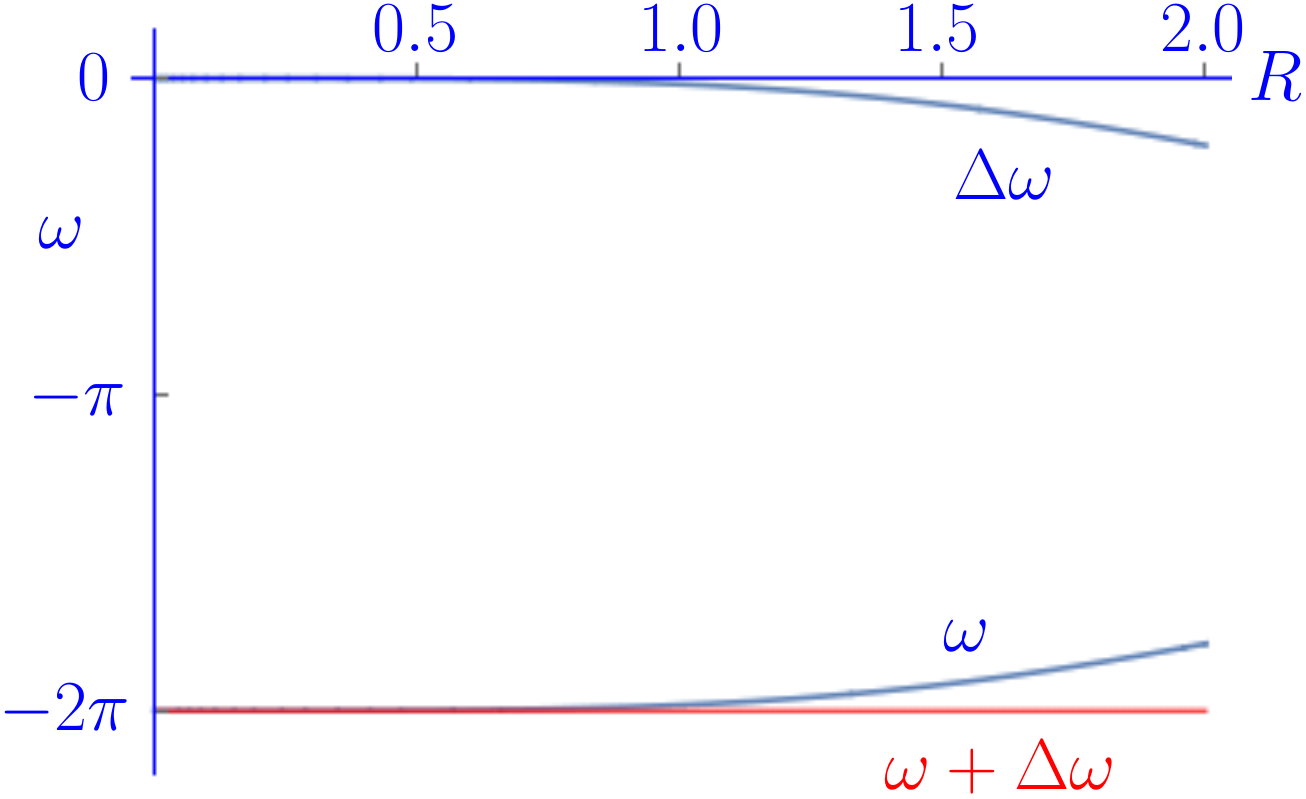}
\caption{Contributions $\omega(R)$ and $\Delta\omega(R)$ to the determination of the rotational angle of the fibre neighbourhood.}
\label{Drehwink}
\end{figure}

As an example of such a configuration in $\mathbb R^3$ with $v=1$ we defined in the diploma thesis~\cite{Jech2014} a field by an area preserving map of $\mathcal{S}^2$ to a circle rotating with $\varphi$ around the line $\rho=\rho_0=3, \zeta=0$, defined in cylindrical coordinates $\rho,\varphi,\zeta$ by
\begin{eqnarray}\begin{aligned}\label{eq:VF_s2winkel}
&\cos\theta=\begin{cases}\frac{(\rho-\rho_0)^2+\zeta^2}{2}-1\le1\\
1\quad\textrm{else}
\end{cases}\\
&\phi=\varphi+\arctan\frac{\zeta}{\rho-\rho_0}.
\end{aligned}\end{eqnarray}
For a given $\vec n$-field one can get the linking number in $\mathbb R^3$ by the famous formula of Carl Friedrich Gauß
\begin{eqnarray}\label{GaussDoppelInt}
v\,=\,\frac{1}{4\pi}\oint_{\mathcal C_1}\oint_{\mathcal C_2}
\frac{\mathbf r_1-\mathbf r_2}{|\mathbf r_1-\mathbf r_2|^3}
\cdot(\mathrm d\mathbf r_1\times\mathrm d\mathbf r_2),
\end{eqnarray}
by a double integral. Another determination of $v$, by a single integral, we get from the observation of the neighbourhood of a single fibre $\mathcal F$. The position along the fibre we parametrize with some parameter $s$. We determine the tangential vector $\mathbf e_\mathcal{F}(s)$ to the fibre and a unit vector to some neighbouring fibre, e.g. defined by the perpendicular component $\mathbf e_\theta(s):=\left.\frac{\pmb\nabla\cos\theta_\mathcal{F}}{|\pmb\nabla\cos\theta_\mathcal{F}|}\right|_\perp$ of the gradient $\pmb\nabla\cos\theta_\mathcal{F}(s)$. We get a local Dreibein with $\mathbf e_\perp(s):=\mathbf e_\mathcal{F}(s)\times\mathbf e_\theta(s)$. Then we determine the rotational velocity along the fibre by
\begin{eqnarray}\label{DiffRot}
\frac{\mathrm d\omega}{\mathrm ds}:=
\mathbf e_\perp\cdot\pmb\nabla_s\mathbf e_\theta
=\mathbf e_\mathcal{F}(\mathbf e_\theta\times\pmb\nabla_s\mathbf e_\theta).
\end{eqnarray}
It turns out that the integrated rotational angle $\omega$ depends on the radius $R=\sqrt{(\rho-\rho_0)^2+\zeta^2}$ of the torus surrounded by the fibre
\begin{eqnarray}\label{LaengsRot}
\omega(R):=\oint_{\mathcal F}\frac{\mathrm d\omega}{\mathrm ds}\mathrm ds\ne2\pi n,\quad n\in\mathcal N
\end{eqnarray}
and is not an integer. Its dependence on $R$ is depicted in Fig.~\ref{Drehwink}. This is understandable from the Gauß map $\mathbf e_\mathcal{F}(s)\mapsto\mathbf e_\mathcal{F}^\prime(s)$ and $\mathbf e_\theta(s)\mapsto\mathbf e_\theta^\prime(s)$ to the $\mathcal{S}_\mathrm{G}^2$ unit sphere, see Fig.~\ref{DrehFaserUmg}. $\mathbf e_\mathcal{F}^\prime(s)$, plotted from the origin of $\mathcal{S}_\mathrm{G}^2$ by $\vec e_\mathcal{F}^{\,\prime}(s)=\vec e_\mathcal{F}(s)$, draws a curve $\mathcal C_\mathcal{F}$ on $\mathcal{S}_\mathrm{G}^2$. \footnote{With bold symbols $\mathbf e$ we indicate the vectors and with $\vec e$ the set of its coordinates.} $\mathbf e_\theta^\prime(s)$ and $\mathbf e_\perp^\prime(s)$ are defined as vectors in the tangential space at $\mathbf e_\mathcal{F}(s)$ by the coordinate equalities $\vec  e_\theta^{\,\prime}(s)=\vec  e_\theta(s)$ and $\vec  e_\perp^{\,\prime}(s)=\vec  e_\perp(s)$. Their parallel transport along $\mathcal C_\mathcal{F}$ is non-trivial, except $\mathcal C_\mathcal{F}$ is a great circle. Deviations from a great circle lead to an additional contribution of the curved geometry on $\mathcal{S}_\mathrm{G}^2$ to the rotational angle. Indicating the position on $\mathbb S_\mathrm{G}^2$ in spherical coordinates $\vec e_\mathcal{F}^{\,\prime}=(\sin\alpha\cos\beta,\sin\alpha\sin\beta,\cos\alpha)$ we can derive for the corresponding spherical coordinate basis $\mathbf e_\alpha$ and $\mathbf e_\beta$  the affine U(1) connection
\begin{eqnarray}\label{AffS2}
C_\alpha=0,\quad C_\beta=\frac{\cos\alpha}{\sin\alpha},\quad
\vec C=C_\alpha\vec e_\alpha+C_\beta\vec e_\beta\hspace{10mm}
\end{eqnarray}
which is trivial for meridians and the equator of $\mathcal{S}_\mathrm{G}^2$. The additional contribution reads
\begin{eqnarray}\label{AddRot}
  \Delta\omega(R):=\oint\mathrm ds\;
  \frac{\mathrm d\vec e_\mathcal{F}^{\,\prime}}{\mathrm ds}\cdot\vec C
\end{eqnarray}
The sum of both contributions gives then the expected value
\begin{eqnarray}\label{AddRot}
\omega(R)+\Delta\omega(R)=2\pi\,v.
\end{eqnarray}
Both contributions of our example are shown in Fig.~\ref{Drehwink}, they give $v=-1$.


\begin{figure}[h!]
\includegraphics[scale=0.6]{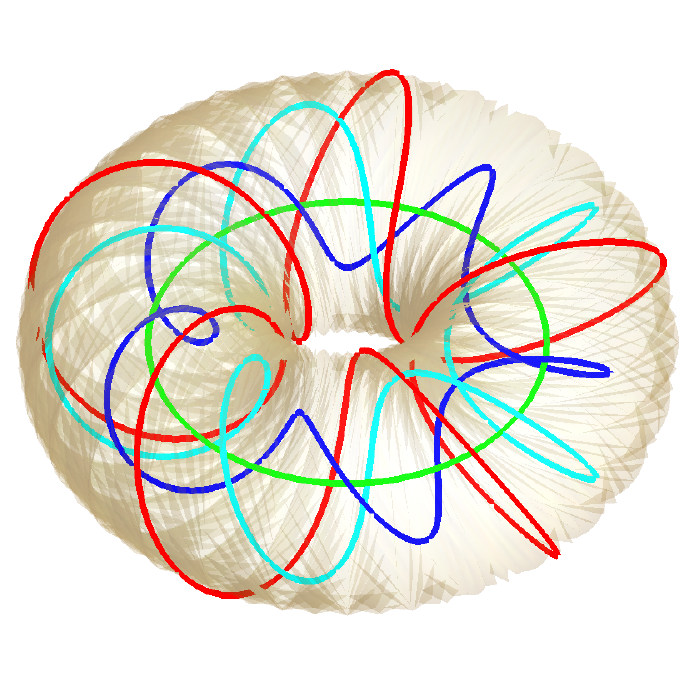}
\includegraphics[scale=0.5]{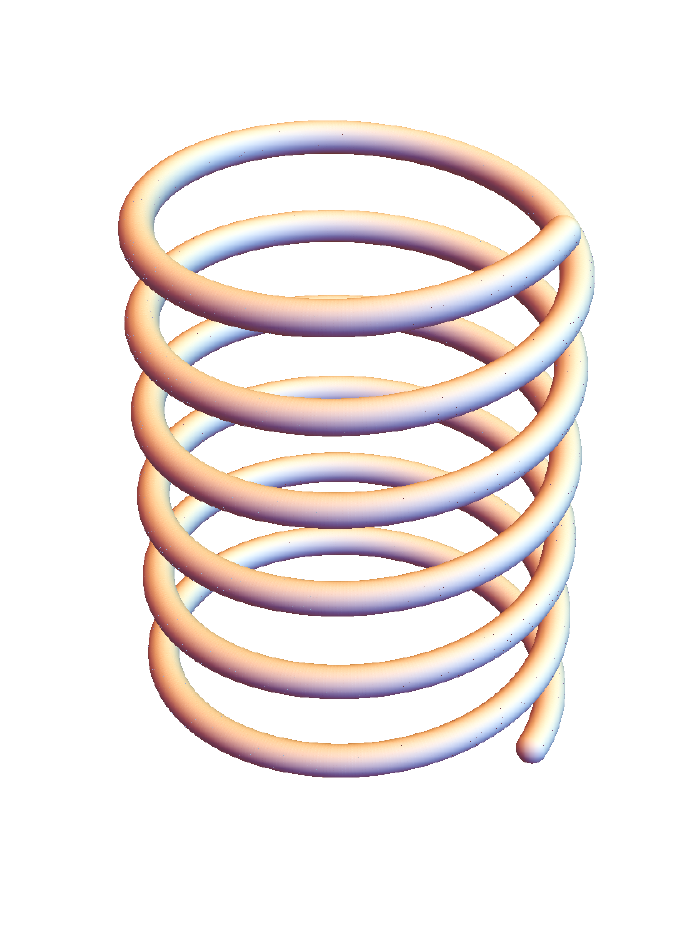}
\caption{Configurations with higher linking numbers arranged with the shape of a torus or in spiral form.}
\label{MehrWind}
\end{figure}

With the conjecture that the photon number of a configuration is given by the Gauß linking number $v$ of fibres, configurations with higher linking numbers $v$ correspond to states with several photons. In such configurations, see Fig.~\ref{MehrWind}, the linked fibres are spiraling several times. The spiral in the right diagram of Fig.~\ref{MehrWind} reminds to circular polarised waves.
\begin{figure}[h!]
\includegraphics[scale=0.3]{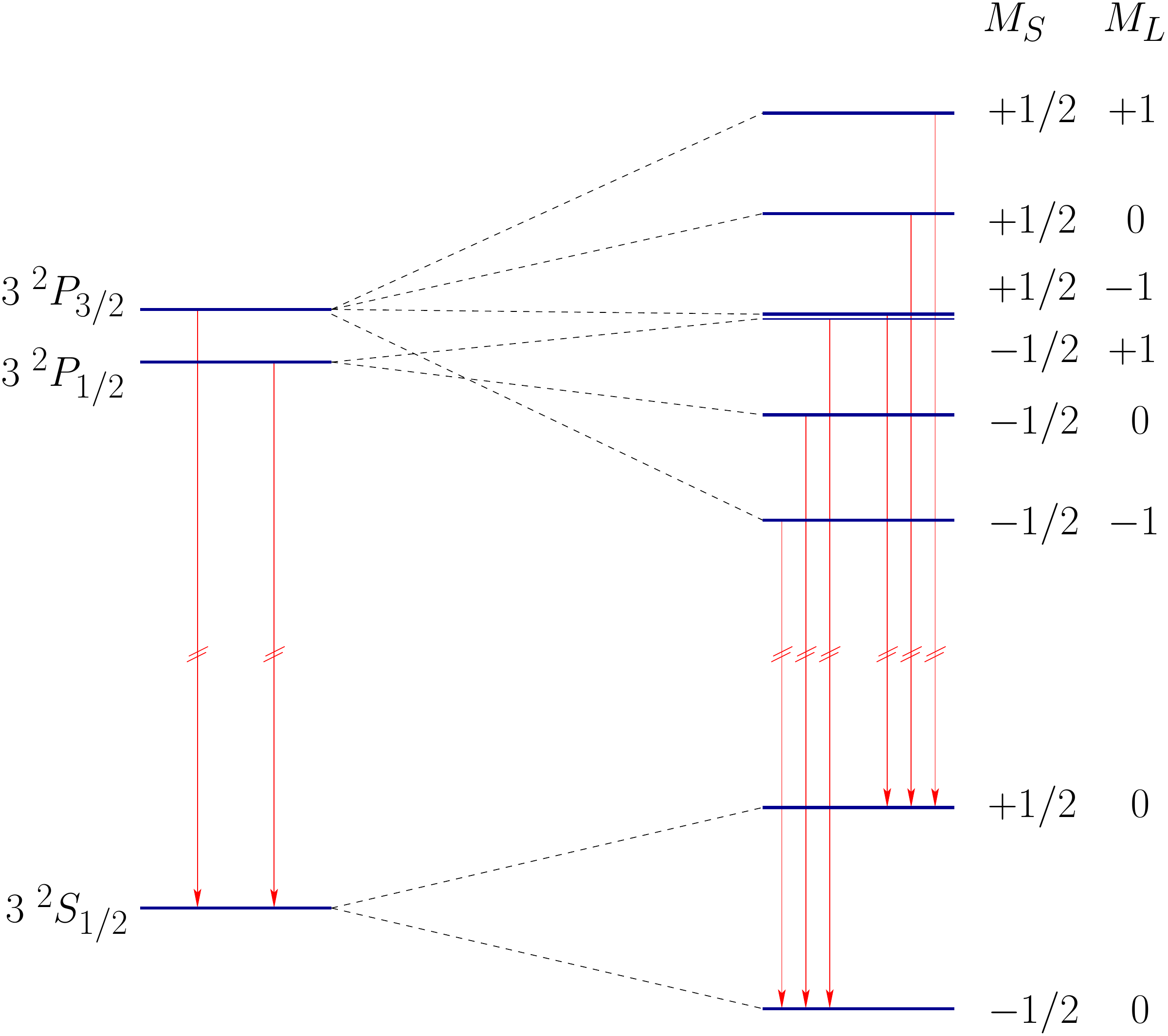}\\
\caption{In a strong magnetic field the Na-D-line splits into three spectral lines with $\Delta M_L=0,\pm1$ and $\Delta M_S=0$.}
\label{LorentzTripel}
\end{figure}
\begin{figure}[h!]
\includegraphics[scale=0.35]{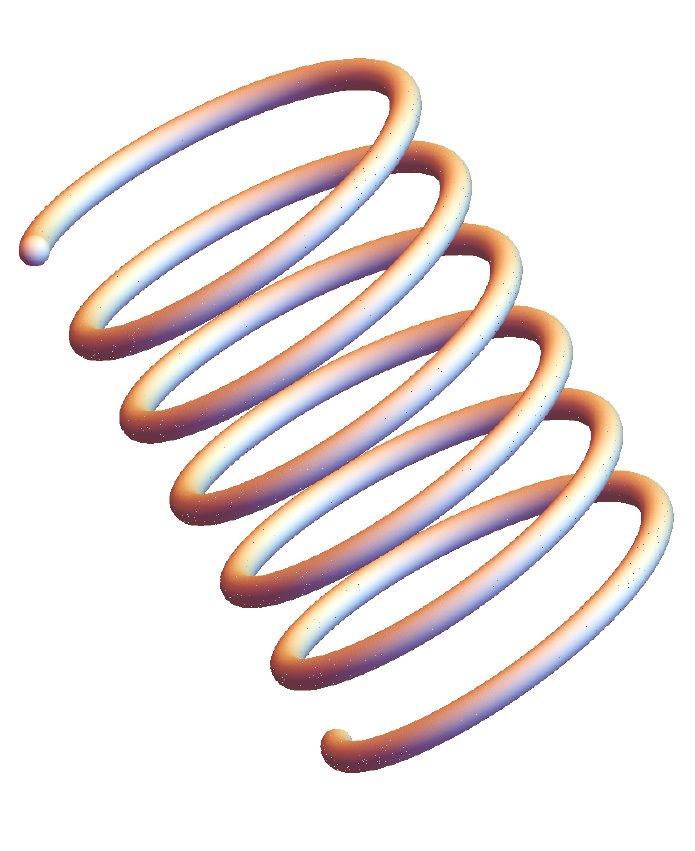}
\includegraphics[scale=0.35]{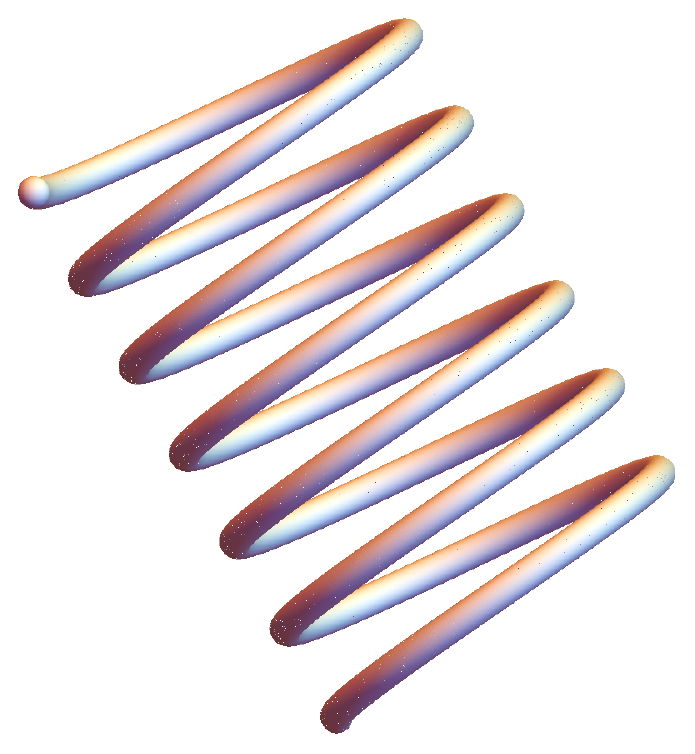}
\includegraphics[scale=0.35]{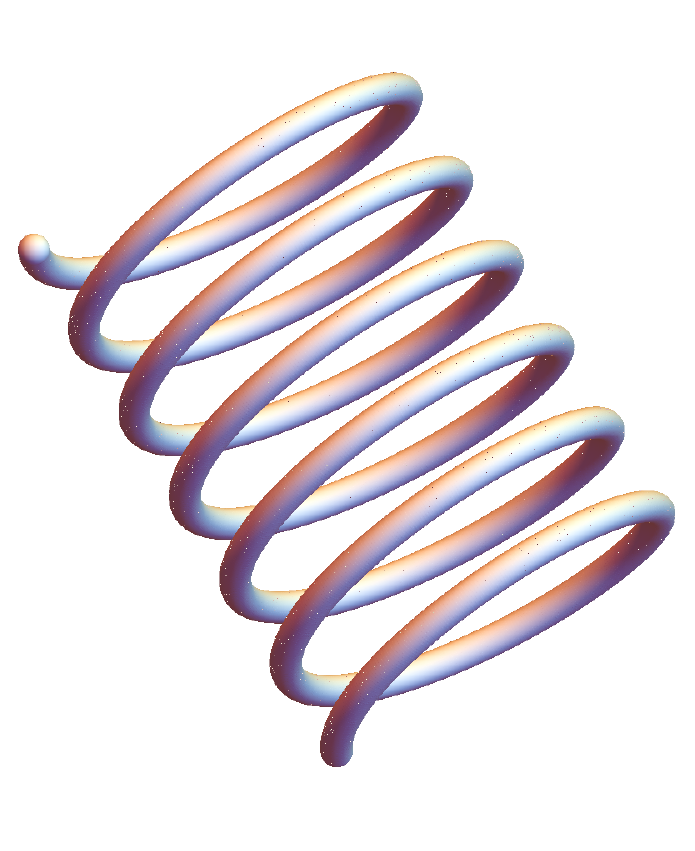}
\includegraphics[scale=0.6]{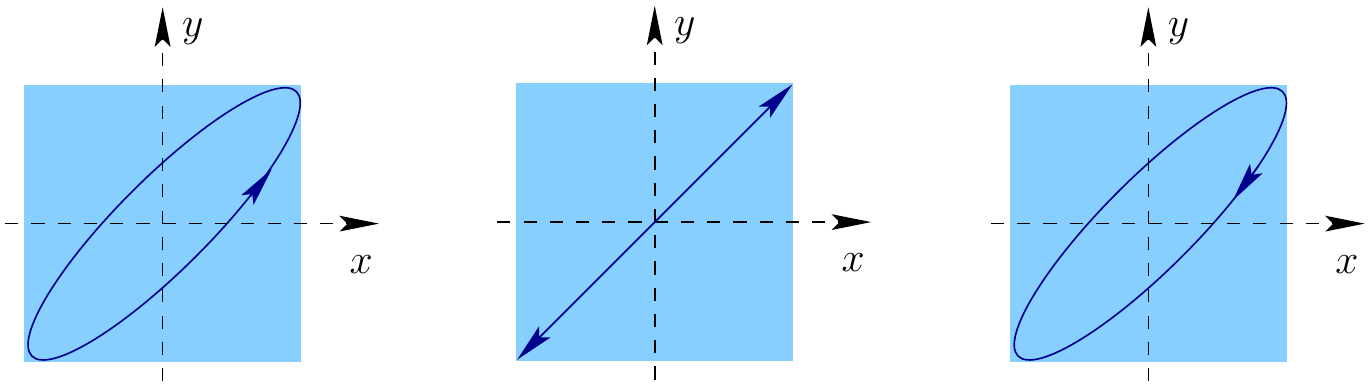}
\caption{Different projections of rotations around a circle produce different polarisations.}
\label{Spiralen}
\end{figure}

\begin{widetext}
\begin{figure}[h!]
\centering
\includegraphics[scale=0.4]{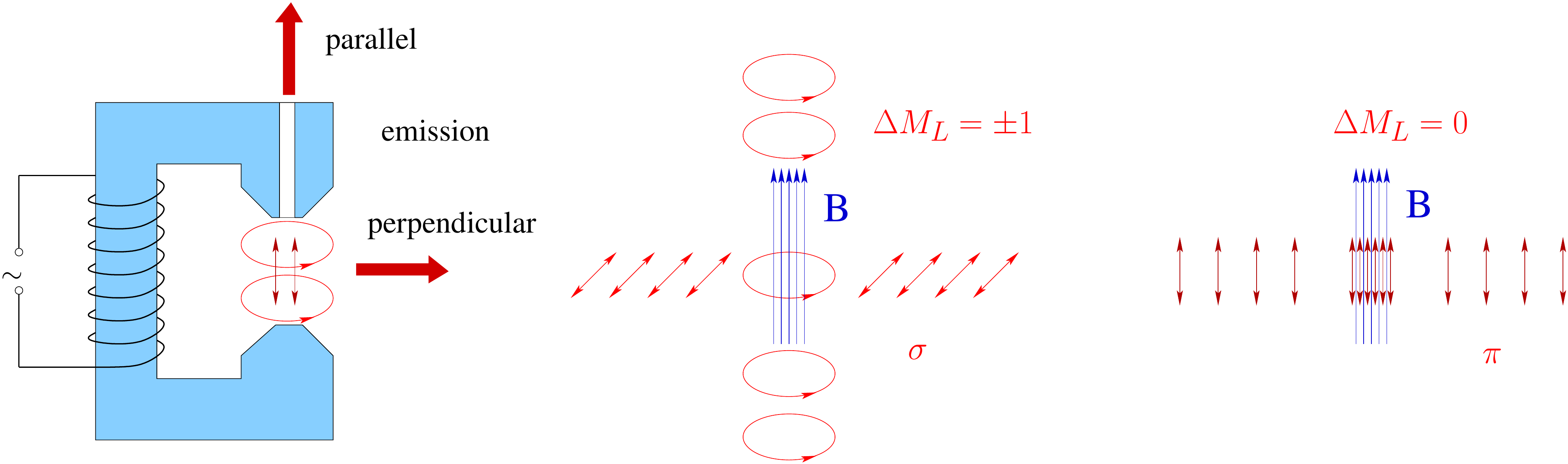}\\
\caption{Scheme of observation of the Na-D-line in a strong magnetic field.}
\label{LinZirkExp}
\end{figure}
\end{widetext}

It may be helpful to have a look at experiments. The Na-D-line in a strong magnetic field splits into three lines, the Lorentz triple, see Fig.~\ref{LorentzTripel}, according to $\Delta M_L=0,\pm1$.
It is interesting to observe the three spectral lines parallel and perpendicular to the magnetic field, as depicted in Fig.~\ref{LinZirkExp}. The lines with $\Delta M_L=0$, the ``$\pi$-components'', are linear polarised with an azimuth dependency of the intensity $I(\vartheta)=\sin^2\vartheta$. The $\Delta M_L=\pm1$-lines, the ``$\sigma$-components'', are circular polarised with $I(\vartheta)=\frac{1}{2}(1+\cos^2\vartheta)$. If the circular polarised lines are observed in the direction perpendicular to the $\mathbf B$-field they appear linearly polarised. This reminds of different projections of circular motion, of a spiral in the comoving x,y,z-ct frame, see Fig.~\ref{Spiralen}.

Here a problem may appear. Rotations of a spiral do not change their chirality. After a rotation by $\pi$ a right-handed spiral remains right-handed as one can easily check with a right-handed thread of a screw. It is well-known that right polarised light can be easily transformed to linear polarised and to left polarised light by quarter-wave plates. These plates have perpendicular fast and slow axes.

This looks as a counter-argument for the interpretation of the Hopf number as photon number. But it may also be an unknown feature of polarised light? Looking carefully at polarisation filters, one realises that filters exist for linear polarised light only! Right and left circular polarised light is detected by transforming it with quarter-wave-plates, by interaction with matter, to linear polarisation.

\section{Conclusion}
The model described in this article is based on a field of SO(3)-matrices in 3+1D Minkowski space-time. Therefore, the model has 3 degrees of freedom only, corresponding to the 3 Euler angles. These rotational matrices can be interpreted as describing the field of orientations of spatial Dreibeins at the points of space-time. In this sense this simple model needs only the degrees of freedom of space and time to describe the various phenomena discussed in this article. To formulate the algebra of this model we are using the simpler representation of rotations by SU(2)-matrices. As a consequence of this treatment we have to take into account that for every SO(3)-configuration there exist two SU(2)-configurations representing the same field. Vector fields (potentials) and tensor fields (field strength) are derived from the basic soliton field $Q(x)$.

The Lagrangian~(\ref{DefL}) of the model contains two terms. The kinetic term is proportional to the square of the curvature $\vec R_{\mu\nu}$ or to the dual field strength. Such a term is well-known from electrodynamics and QCD. The potential term reminds of the Higgs potential and the cosmological constant. According to the Hobart-Derrick theorem a quarter of the mass of solitons would contribute with its average to the cosmological constant. The transition from $Q=1\textrm{ and }\Lambda=1/r_0^4$ to $\Lambda=0$ releases an energy density of
$\alpha_f\hbar c/(4\pi r_0^4)=4.8~\textrm{keV/fm}^3=7.7\cdot 10^{29}J/\textrm m^3$ and could have contributed to inflation. Further, the potential term allows for a two-dimensional degeneracy of the vacuum states.

In this model there appear particle-like excitation as topological solitons. Their mass is field energy only. The field configurations are characterised by three topological quantum numbers which could find a physical realisation as electric charge, spin quantum number and photon number. The topological structure does not allow two solitons to occupy the same space. This could be the topological origin of the Pauli principle. Charges appear as integer multiples of an elementary charge. No fractional charges are possible for stable excitations. Charges are characterised by regions where the spatial Dreibein rotates by $2\pi$. Such regions interact via Coulomb and Lorentz forces. The electric force follows the $1/r^2$ behaviour at large distances and increases for high momentum transfers. The distinction between charges and their fields is unnecessary. Such a distinction can be introduced as an approximation and simplifies the comparison with our physical experience. Eigen-angular momentum (spin) appears as a consequence of orbital motion. The degenerate vacuum allows for two types of massless excitations, which could be related to the two polarisations of photons. U(1) gauge invariance emerges in the electrodynamic limit as rotational invariance of the above mentioned Dreibein around the $\vec n$-axis.

Despite the small number of degrees of freedom there are further non-quantized  disturbances possible. Magnetic currents as violations of the homogeneous Maxwell equations propagating with the speed of light which contribute to force fields only via their field strengths. Further we find waves in the rotational angle $\omega=2\alpha$ with energy contributions from the potential term in the Hamiltonian. The cosmological discoveries of recent years allow to attribute these two types of disturbances to dark energy and dark matter contributions.

The investigations seem to support the conjecture that the particles we find in experiments are topological solitons characterised by topological quantum numbers. They produce only spots on films and clicks in detectors. Waves seem to escape and are not directly detected. Waves could disturb the paths of particles and could contribute to a subquantum medium leading to quantum mechanics. In analogy to Couder's silicon oil drop experiments~\cite{PhysRevLett.94.177801,PhysRevLett.97.154101,PhysRevLett.102.240401,2013arXiv1307.6051F} the interference of waves created by the particle themselves with the vacuum waves could serve as a guiding wave field for particles.

If this model reproduced some features of nature the two types of long range forces which we find, gravitational and Coulomb forces would be described in a geometrical manner with the degrees of freedom of space-time only. This would give a hint that particle physics could be closely related to gravitation.

\subsection*{Conflicts of Interest}
The author declares that there are no conflicts of interest
regarding the publication of this paper.

\bibliographystyle{unsrt}
\bibliography{literatur}

\begin{thebibliography}{10}

\bibitem{Remoissenet:1999wa}
Michel Remoissenet.
\newblock {\em Waves Called Solitons: Concepts and Experiments}.
\newblock Advanced Texts in Physics. Springer, 2003.

\bibitem{misner1973gravitation}
C.W. Misner, K.S. Thorne, and J.A. Wheeler.
\newblock {\em Gravitation}.
\newblock Number Teil 3 in Gravitation. W. H. Freeman, 1973.

\bibitem{Faber:1999ia}
Manfried Faber.
\newblock A model for topological fermions.
\newblock {\em Few Body Syst.}, 30:149--186, 2001.

\bibitem{Faber:2002nw}
Manfried Faber and Alexander~P. Kobushkin.
\newblock {Electrodynamic limit in a model for charged solitons}.
\newblock {\em Phys.Rev.}, D69:116002, 2004.

\bibitem{Borisyuk:2007bd}
Dmitry Borisyuk, Manfried Faber, and Alexander Kobushkin.
\newblock {Electro-Magnetic Waves within a Model for Charged Solitons}.
\newblock {\em J.Phys.}, A40:525--531, 2007.

\bibitem{Faber:2008hr}
Manfried Faber, Alexander Kobushkin, and Mario Pitschmann.
\newblock {Shape vibrations of topological fermions}.
\newblock {\em Adv.Stud.Theor.Phys.}, 2:11--22, 2008.

\bibitem{Faber:2012zz}
Manfried Faber.
\newblock {Particles as stable topological solitons}.
\newblock {\em J.Phys.Conf.Ser.}, 361:012022, 2012.

\bibitem{Faber:2014bxa}
Manfried Faber.
\newblock {Spin and charge from space and time}.
\newblock {\em J.Phys.Conf.Ser.}, 504:012010, 2014.

\bibitem{Rodrigues:1840aa}
B.O. Rodrigues.
\newblock {Des lois géométriques qui régissent les déplacements d'un
  système solide dans l'espace, et de la variation des coordonnées provenant
  de ces déplacements considérés indépendamment des causes qui peuvent les
  produire}.
\newblock {\em Journal de Mathématiques Pures et Appliquées}, 5:380--440,
  1840.

\bibitem{Skyrme:1961vq}
T.H.R. Skyrme.
\newblock {A Nonlinear field theory}.
\newblock {\em Proc. Roy. Soc. Lond.}, A260:127--138, 1961.

\bibitem{Hobart:1963rh}
R.H. Hobart.
\newblock On the instability of a class of unitary field models.
\newblock {\em Proc. Phys. Soc. London}, 82(2):201--203, 1963.

\bibitem{Derrick:1964gh}
G.H. Derrick.
\newblock Comments on nonlinear wave equations as models for elementary
  particles.
\newblock {\em J. Math. Phys.}, 5(2):1252--1254, 1964.

\bibitem{jackson_classical_1999}
John~David Jackson.
\newblock {\em Classical electrodynamics}.
\newblock Wiley, New York, {NY}, 3rd ed. edition, 1999.

\bibitem{Wabnig2001}
Joachim Wabnig.
\newblock {Interaction in the model of topological fermions}.
\newblock Master's thesis, Techn. Univ. Wien, 2001.

\bibitem{Resch2011}
Josef Resch.
\newblock {Numerische Analyse an Dipolkonfigurationen im Modell topologischer
  Fermionen}.
\newblock Master's thesis, Techn. Univ. Wien, 2011.

\bibitem{Theuerkauf2016}
Dominik Theuerkauf.
\newblock {Charged particles in the model of topological fermions}.
\newblock Master's thesis, Techn. Univ. Wien, 2016.

\bibitem{Wu:1975vq}
T.~T Wu and Chen-Ning Yang.
\newblock Some remarks about unquantized nonabelian gauge fields.
\newblock {\em Phys. Rev.}, D12:3843--3844, 1975.

\bibitem{Faddeev:1975tz}
L.~D. Faddeev.
\newblock {Quantization of Solitons}.
\newblock In {\em {Tbilisi Conf.1976:0T50}}, 1975.

\bibitem{Faddeev1976}
L.~D. Faddeev.
\newblock Some comments on the many-dimensional solitons.
\newblock {\em Letters in Mathematical Physics}, 1(4):289--293, 1976.

\bibitem{Ferreira:2000hz}
L.~A. Ferreira and A.~V. Razumov.
\newblock {Hopf solitons and area preserving diffeomorphisms of the sphere}.
\newblock {\em Lett. Math. Phys.}, 55:143--148, 2001.

\bibitem{Ferreira:2006wg}
L.~A. Ferreira.
\newblock {Exact time dependent Hopf solitons in 3+1 dimensions}.
\newblock {\em JHEP}, 03:075, 2006.

\bibitem{Jech2014}
Markus Jech.
\newblock {Die Hopfzahl in einer SU(2)-Feldtheorie}.
\newblock Master's thesis, Techn. Univ. Wien, 2014.

\bibitem{PhysRevLett.94.177801}
Y.~Couder, E.~Fort, C.-H. Gautier, and A.~Boudaoud.
\newblock From bouncing to floating: Noncoalescence of drops on a fluid bath.
\newblock {\em Phys. Rev. Lett.}, 94:177801, May 2005.

\bibitem{PhysRevLett.97.154101}
Yves Couder and Emmanuel Fort.
\newblock Single-particle diffraction and interference at a macroscopic scale.
\newblock {\em Phys. Rev. Lett.}, 97:154101, Oct 2006.

\bibitem{PhysRevLett.102.240401}
A.~Eddi, E.~Fort, F.~Moisy, and Y.~Couder.
\newblock Unpredictable tunneling of a classical wave-particle association.
\newblock {\em Phys. Rev. Lett.}, 102:240401, Jun 2009.

\bibitem{2013arXiv1307.6051F}
E.~{Fort}, A.~{Eddi}, A.~{Boudaoud}, J.~{Moukhtar}, and Y.~{Couder}.
\newblock {Path-memory induced quantization of classical orbits}.
\newblock {\em ArXiv e-prints}, July 2013.

\end{thebibliography}
\end{document}